\documentclass[12pt]{elsart}
\usepackage{epsfig}
\def\boldsymbol#1{\mbox{\boldmath ${#1}$}}

\begin{document}

\begin{frontmatter}

\title{A measurement of the axial form factor of the nucleon\\
by the $\mathrm{p}(\mathrm{e},\mathrm{e}'\pi^+)\mathrm{n}$
reaction at $W=1125\,\mathrm{MeV}$}

\author[KPH,PHD]{A.~Liesenfeld},
\author[KPH,PHD]{A.~W.~Richter}, 
\author[IJS,PHD,COR]{S. \v Sirca},
\author[KPH]{K.~I.~Blomqvist}, 
\author[KPH]{W.~U.~Boeglin}, 
\author[IJS]{K.~Bohinc},
\author[KPH]{R.~B\"ohm},
\author[KPH]{M.~Distler}, 
\author[KPH]{D.~Drechsel},
\author[KPH]{R.~Edelhoff},
\author[KPH]{I.~Ewald},
\author[KPH]{J.~Friedrich},
\author[KPH]{J.~M.~Friedrich},
\author[KPH]{R.~Geiges},
\author[KPH]{M.~Kahrau}, 
\author[KPH]{M.~Korn}, 
\author[KPH]{K.~W.~Krygier}, 
\author[KPH]{V.~Kunde}, 
\author[KPH]{H.~Merkel}, 
\author[KPH]{K.~Merle}, 
\author[KPH]{U.~M\"uller}, 
\author[KPH]{R.~Neuhausen}, 
\author[KPH]{T.~Pospischil}, 
\author[IJS]{M.~Potokar}, 
\author[IJS]{A.~Rokavec}, 
\author[KPH]{G.~Rosner}, 
\author[KPH]{P.~Sauer}, 
\author[KPH]{S.~Schardt}, 
\author[KPH]{H.~Schmieden},
\author[KPH]{L.~Tiator},
\author[IJS]{B.~Vodenik},
\author[KPH]{A.~Wagner}, 
\author[KPH]{Th.~Walcher} and 
\author[KPH]{S.~Wolf}

\address[KPH]{Institut f\"ur Kernphysik, Universit\"at Mainz,
              D-55099 Mainz, Germany}
\address[IJS]{Jo\v zef Stefan Institute,
              SI-1001 Ljubljana, Slovenia}

\thanks[PHD]{This paper comprises parts of the doctoral theses of
             A.~Liesenfeld, A.~W.~Richter and S.~\v Sirca.}
\thanks[COR]{Corresponding author (tel: +386 61 1773-731,
                                   fax: +386 61 219-385,\\
                                   e-mail: simon.sirca@ijs.si).}

\date{\today}

\begin{abstract}
The reaction $\mathrm{p}(\mathrm{e},\mathrm{e}'\pi^+)\mathrm{n}$
was measured at the Mainz Microtron MAMI at an invariant mass
of $W=1125\,\mathrm{MeV}$ and four-momentum transfers
of $Q^2=0.117$, $0.195$ and $0.273\,(\mathrm{GeV/c})^2$.
For each value of $Q^2$, a Rosenbluth separation
of the transverse and longitudinal cross sections was performed.
An effective Lagrangian model was used to extract the `axial mass'
from experimental data.  We find a value of
$M_{\mathrm{A}}=(1.077\pm 0.039)\,\mathrm{GeV}$
which is $(0.051\pm 0.044)\,\mathrm{GeV}$ larger than the
axial mass known from neutrino scattering experiments.
This is consistent with recent calculations
in chiral perturbation theory.
\end{abstract}

\end{frontmatter}

{\it PACS:} 13.60.Le, 25.30.Rw, 14.20.Dh

{\it Keywords:} nucleon axial form factor,
                coincident pion electroproduction

\newpage

\section{Introduction}

There are basically two methods to determine the weak axial
form factor of the nucleon.  One set of experimental data comes
from measurements of (quasi)elastic (anti)neutrino scattering
on protons \cite{fanourakisX}, deuterons \cite{barishX}
and other nuclei (Al, Fe) \cite{holder,kustom}
or composite targets like freon \cite{perkins,orkin,bonetti,armenise}
and propane \cite{armenise,budagov}.  In the (quasi)elastic picture of
(anti)neutrino-nucleus scattering, the $\nu\mathrm{N}\to\mu\mathrm{N}$
weak transition amplitude can be expressed in terms of
the nucleon electromagnetic form factors $F_1$ and $F_2$ and the axial
form factor $G_{\mathrm{A}}$.  The axial form factor is then extracted
by fitting the $Q^2$-dependence of the (anti)neutrino-nucleon
cross section,
\begin{equation}
{{\d}\sigma\over{\d}Q^2}=
A(Q^2) \mp B(Q^2)\,(s-u) + C(Q^2)\,(s-u)^2\>{,}
\end{equation}
in which $G_{\mathrm{A}}(Q^2)$ is contained in the bilinear forms
$A(Q^2)$, $B(Q^2)$ and $C(Q^2)$ of the relevant form factors
and is assumed to be the only unknown quantity.
It can be parameterised in terms of an `axial mass' $M_{\mathrm{A}}$
as ${G_{\mathrm{A}}(Q^2) =
G_{\mathrm{A}}(0)}/(1 + Q^2/M_{\mathrm{A}}^2)^2$.

\begin{figure}[h]
\begin{center}
\includegraphics[height=7cm]{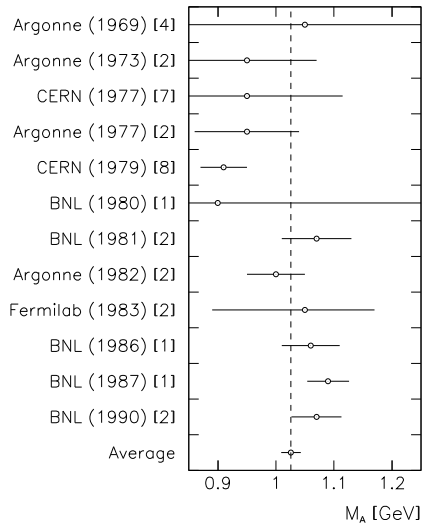}
\end{center}
\vspace*{-5mm}
\caption{Axial mass $M_{\mathrm{A}}$ as extracted from
(quasi)elastic neutrino and antineutrino scattering experiments.
The weighted average is
$M_{\mathrm{A}}=(1.026\pm 0.017)\,\mathrm{GeV}$, or
$(1.026\pm 0.021)\,\mathrm{GeV}$ using the scaled-error
averaging recommended by Ref.~\protect\cite{PDG}.}
\label{fig:m_a_nu}
\end{figure}

Fig.~\ref{fig:m_a_nu} shows the available values for
$M_{\mathrm{A}}$ obtained from these studies.
References \cite{holder,perkins,orkin,budagov} reported severe
uncertainties in either knowledge of the incident neutrino flux or
reliability of the theoretical input needed to subtract the background
from genuine elastic events (both of which gradually improved in
subsequent experiments).  The values derived fall well outside
the most probable range of values known today and exhibit very large
statistical and systematical errors.  Following the data selection
criteria of the Particle Data Group \cite{PDG},
they were excluded from this compilation.

Another body of data comes from charged pion electroproduction
on protons \cite{esau,olsson,amaldiX,joos,choi,nambu,bloo}
slightly above the pion production threshold.
As opposed to neutrino scattering, which is described
by the Cabibbo-mixed $V-A$ theory, the extraction of the axial
form factor from electroproduction requires a more involved
theoretical picture.

\begin{figure}[ht]
\begin{center}
\includegraphics[height=9.5cm]{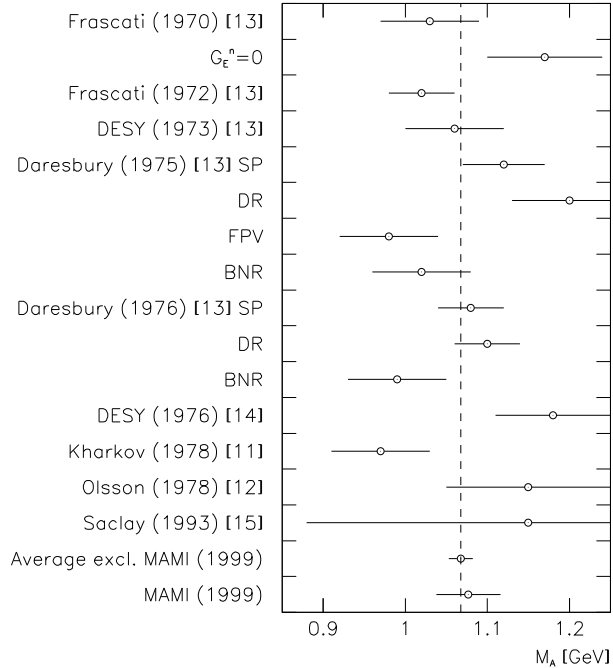}
\end{center}
\vspace*{-5mm}
\caption{Axial mass $M_{\mathrm{A}}$ as extracted from
charged pion electroproduction experiments.
The weighted average (excluding our result)
is $M_{\mathrm{A}}=(1.068\pm 0.015)\,\mathrm{GeV}$,
or $(1.068\pm 0.017)\,\mathrm{GeV}$ using the scaled-error
averaging \protect\cite{PDG}.  Including our
extracted value, the weighted scaled-error average becomes
$M_{\mathrm{A}}=(1.069\pm 0.016)\,\mathrm{GeV}$.
Note that our value contains both the statistical {\sl and\/}
systematical uncertainty; for other values the systematical
errors were not explicitly given.
SP:~soft-pion limit, DR:~analysis using approach of
Ref.~\protect\cite{DR}, FPV:~Ref.~\protect\cite{FPV},
BNR:~Ref.~\protect\cite{BNR}.}
\label{fig:m_a_ee}
\end{figure}

The basic result about low energy photoproduction
of massless charged pions can be traced back to the Kroll-Ruderman
theorem \cite{KR}, extended to virtual photons by Nambu, Luri\'e
and Shrauner \cite{NLS}, who obtained the $\mathcal{O}(Q^2)$ result
for the isospin $(-)$ (see Ref.~\cite{xs}, p.~29 for notation)
electric dipole amplitude at threshold
\begin{equation}
E_{0+}^{(-)}(m_\pi=0,Q^2)={\mathrm{e}g_{\mathrm{A}}\over
8\pi f_\pi}\,\biggl\{ 1 - {Q^2\over 6}\,\langle
r_{\mathrm{A}}^2\rangle - {Q^2\over 4M^2}\,\biggl[\,
\kappa_{\mathrm{v}} + {1\over 2}\,\biggr]
+ \mathcal{O}(Q^3) \biggr\}\>{,}
\label{eq:NLS}
\end{equation}
where $\kappa_{\mathrm{v}}$ is the nucleon isovector anomalous
magnetic moment, $g_{\mathrm{A}}\equiv G_{\mathrm{A}}(0)$
is the axial coupling constant,
and $f_\pi$ is the pion decay constant.

In the following years, improved models were proposed
\cite{nambu,FPV,DR,BNR}, most of them including corrections
due to the finite pion mass $\mu=m_\pi/M$.
The values of the axial mass were determined,
within the framework of the respective model,
from the slopes of the angle-integrated differential
electroproduction cross sections at threshold,
\begin{equation}
\langle r_{\mathrm{A}}^2\rangle =
-{6\over G_{\mathrm{A}}(0)}\,
{{\d}G_{\mathrm{A}}(Q^2)\over{\d}Q^2}\,
\biggl\vert_{Q^2=0} = {12\over M_{\mathrm{A}}^2}\>{.}
\end{equation}
The results of various measurements and theoretical approaches
are shown in Fig.~\ref{fig:m_a_ee}.  Note again that references
\cite{nambu,bloo} were omitted from the fit for lack of reasonable
compatibility with the other results.

Although the results of these investigations deviate from
each other by more than their claimed accuracy, the weighted
averages from neutrino scattering and electroproduction
give quite precise values of the axial mass.
Comparing the average values
of the two methods, one observes a significant difference of
$\Delta M_{\mathrm{A}}=(0.042\pm 0.023)\,\mathrm{GeV}$, or
$(0.042\pm 0.027)\,\mathrm{GeV}$ using the scaled-error averaging.

Chiral perturbation theory ($\chi\mathrm{PT}$) has recently
shown a remarkable and model-in\-de\-pen\-dent result
that already at $\mathcal{O}(Q^2)$,
the NLS result of eq.~(\ref{eq:NLS}) is strongly
modified due to pion loop contributions \cite{BKM1}.  
These contributions effectively reduce
the mean-square axial radius,
\begin{equation}
\langle r_{\mathrm{A}}^2\rangle \to
\langle r_{\mathrm{A}}^2\rangle 
+ {3\over 64f_\pi^2}\,\biggl( 1 - {12\over\pi^2} \biggr)\>\mathrm{.}
\label{eq:chiPTcorr}
\end{equation}
The loop correction in eq.~(\ref{eq:chiPTcorr}) has a value of
$-0.046\,\mathrm{fm}^2$, which is a $-10\,\%$ correction to a typical
$\langle r_{\mathrm{A}}^2\rangle=0.45\,\mathrm{fm}^2$.
Correspondingly, the axial mass $M_{\mathrm{A}}=
\sqrt{12}/\langle r_{\mathrm{A}}^{2}\rangle^{1/2}$
would appear to be about $5\,\%$ larger in electroproduction
than in neutrino scattering,
in agreement with the observed $\Delta M_{\mathrm{A}}$.
The aim of the present investigation was to determine
$M_{\mathrm{A}}$ from new, high precision pion electroproduction
data and thereby help verify whether this discrepancy was genuine.
Since both the energy and momentum transfers were too high to allow
for a safe extraction of $E_{0+}^{(-)}$, these data
were analysed in the framework of an effective Lagrangian model
with the electromagnetic nucleon form factors, the electric pion
form factor and the axial nucleon form factor
at the appropriate vertices \cite{piplus1,dt}.

\section{Kinematics of the experiment}

The differential cross sections for $\pi^+$ electroproduction
on protons were measured at an invariant mass
of $W=1125\,\mathrm{MeV}$ and at four-momentum transfers
of the virtual photon $Q^2=0.117$, $0.195$ and
$0.273\,(\mathrm{GeV/c})^2$.  For each value of $Q^2$, we measured
the scattered electron and the outgoing pion in parallel kinematics
at three different polarisations of the virtual photon, enabling us
to separate the transverse and the longitudinal part of the
cross section by the Rosenbluth technique.
Table~\ref{tab:settings} shows the experimental settings.

\begin{table}[ht]
\caption{Experimental settings for the
$\mathrm{p}(\mathrm{e},\mathrm{e}'\pi^+)\mathrm{n}$ experiment.
The angles are measured with respect to the electron beam axis.}
\label{tab:settings}
\begin{center}
\vspace*{6pt}
\begin{tabular}{ccccrrc}
\hline

\hline

\hline
Setting ($\varepsilon$) & $Q^2$      & $E_{\mathrm{e}}$
  & $E_{\mathrm{e}}'$   & $\theta_{\mathrm{e}}$
  & $\theta_\pi$        & $p_\pi$ \\[-5pt]
  & $[\mathrm{GeV}^2/\mathrm{c}^2]$  & $[\mathrm{MeV}]$
  & $[\mathrm{MeV}]$    & $[^\circ]$ & $[^\circ]$ 
  & $[\mathrm{MeV/c}]$\\

\hline\\

$0.834$ & $0.117$ & $855.11$ & $587.35$ & $-27.93$ & $39.31$  
        & $188.84$ \\
$0.500$ &         & $510.11$ & $242.35$ & $-58.22$ & $28.31$  
        & $$ \\
$0.219$ &         & $405.11$ & $137.36$ & $92.96$  & $-18.41$
        & $$ \\
$0.742$ & $0.195$ & $855.11$ & $545.79$ & $-37.72$ & $38.27$
        & $209.62$ \\
$0.437$ &         & $585.11$ & $275.89$ & $66.67$  & $-28.03$
        & $$ \\
$0.229$ &         & $495.11$ & $185.79$ & $93.45$  & $-20.12$
        & $$ \\
$0.648$ & $0.273$ & $855.11$ & $504.55$ & $46.83$  & $-35.82$
        & $228.00$ \\
$0.457$ &         & $690.11$ & $339.55$ & $65.27$  & $-29.37$
        & $$ \\
$0.259$ &         & $585.11$ & $234.55$ & $89.60$  & $-21.90$
        & $$ \\
[5pt]
\hline

\hline

\hline
\end{tabular}
\end{center}
\end{table}

\section{Experimental setup}

The measurements were performed at the Institut f\"ur Kernphysik
at the University of Mainz, using the continuous-wave
electron microtron MAMI \cite{mami}.  The energies of the incoming
electron beam ranged from $405$ to $855\,\mathrm{MeV}$,
and the beam energy spread did not exceed $0.16\,\mathrm{MeV}$.
The $15$ to $35\,\mu\mathrm{A}$ electron beam was scattered
on a liquid hydrogen target cell
($5\times 1\times 1\,\mathrm{cm}^3$ with $10\,\mu\mathrm{m}$
Havar walls in settings with the photon polarisation parameter
$\varepsilon = 0.437$, $0.648$, $0.457$ and $0.259$,
and on a $2\,\mathrm{cm}$-diameter cylindrical target cell
with $50\,\mu\mathrm{m}$ Havar walls in all other settings)
attached to a high power target cooling system.
Forced circulation and a beam wobbling system were used
to avoid density fluctuations of the liquid hydrogen.
With this system, luminosities of up to
$3.2\cdot 10^{37}\mathrm{cm}^{-2}\mathrm{s}^{-1}$
($32\,\mathrm{MHz}/\mu\mathrm{b}$) were attained.

\newpage

The scattered electrons and the produced pions were detected in
coincidence by the high resolution ($\delta p/p\approx 10^{-4}$)
magnetic spectrometers A (SpecA) and B (SpecB)
of the A1 Collaboration \cite{neuhaus}.
In settings with $\varepsilon = 0.834$, $0.500$ and $0.742$,
electrons were detected with SpecB and pions with SpecA,
and vice versa in all other settings.
The momentum acceptance $\Delta p/p$ was
$20\,\%$ and $15\,\%$ in SpecA and SpecB, respectively.
Heavy-metal collimators ($21\,\mathrm{msr}$ in SpecA, $5.1$ or
$5.6\,\mathrm{msr}$ in SpecB) were used to minimise angular
acceptance uncertainties due to the relatively large target cells.

A trigger detector system consisting of two planes of segmented
plastic scintillators and a threshold \v Cerenkov detector
were used in each spectrometer.  The coincidence time resolution,
taking into account the different times of flight of particles for
different trajectory lengths through the spectrometer, was between
$1.0$ and $2.6\,\mathrm{ns}$ FWHM.

In each spectrometer, four vertical drift chambers were used for
particle tracking, measurement of momenta and target vertex
reconstruction.  Back-tracing the particle trajectories from
the drift chambers through the magnetic systems, an angular
resolution (all FWHM) better than $\pm 5\,\mathrm{mrad}$
(dispersive and non-dispersive angles) and spatial (vertex)
resolution better than $\pm 5\,\mathrm{mm}$ (non-dispersive
direction, SpecA) and $\pm 1\,\mathrm{mm}$ (SpecB) were achieved
on the target.  A more detailed description of the apparatus
can be found in Ref.~\cite{neuhaus}.

\section{Data analysis}

In the offline analysis, cuts in the corrected coincidence time
spectrum were applied to identify real coincidences and to eliminate
the background of accidental coincidences.
A cut in the energy deposited in the first scintillator plane
in the pion spectrometer was used to discriminate
charged pions against protons.  The \v Cerenkov signal was
used to identify electrons in the electron spectrometer
and to veto against positrons in the pion spectrometer.

The true coincidences were observed in the peak
of the accumulated missing mass distribution
$(E_\mathrm{miss}^2-|\vec{p}_\mathrm{miss}|^2)^{1/2}-M_\mathrm{n}$,
using an event-by-event reconstruction of
$(E_\mathrm{miss},\vec{p}_\mathrm{miss})=(\omega+M_\mathrm{p}-E_\pi,
\vec{q}-\vec{p}_\pi)$.

The cross sections were subsequently corrected for
detector and coincidence inefficiencies
(between $+1.7\,\%$ and $+3.2\,\%$)
and dead-time losses (between $+1.4\,\%$ and $+4.6\,\%$).
The detector efficiencies and their uncertainties were
measured by the three-detector method, while the coincidence
efficiency of the setup and the uncertainty of the dead-time
measurement were determined in simultaneous
single-arm and coincidence measurements of elastic
$\mathrm{p}(\mathrm{e},\mathrm{e}'\mathrm{p})$,
which was also used to check on the acceptance
for the extended targets.

The accepted phase space was determined by a Monte Carlo
simulation which provided the (event-wise) Lorentz transformation
to the CM system and incorporated radiative corrections and
ionisation losses of the incoming and scattered electrons and pions.
Full track was kept of the particles' trajectories
and their lengths in target and detector materials.
Since exact energy losses are not known event-wise,
we used the most probable energy losses
for the subsequent energy loss correction in both simulation
and analysis programs, and compared the corresponding
missing mass spectra.  The uncertainty estimates were based
on relative variations of their content in dependence
of the cut-off energy along the radiative tail.

The uncertainty of the integrated luminosity originates
only in the target density changes due to temperature fluctuations
within the target cell, while the electron beam current
is virtually exactly known.
Finally, a computer simulation was used to determine
the correction factors due to the pion decaying in flight
from the interaction point to the scintillation detectors,
taking into account the muon contamination at the target
(correction factors ranging from $\times 2.23$ to $\times 2.89$
in different settings).  The systematical errors of the
pion decay correction factors were estimated
from the statistical fluctuations of the back-traced
muon contamination at the target.

\section{Results and discussion}

In the Born approximation, the coincidence cross section
for pion electroproduction can be factorised as \cite{piplus1,xs}
\begin{equation}
{{\d}\sigma\over{\d}E'_{\mathrm{e}}\,
{\d}\Omega_{\mathrm{e}}'\,{\d}\Omega_\pi^\star} = 
\Gamma_{\mathrm{v}}
{{\d}\sigma_{\mathrm{v}}\over
{\d}\Omega_\pi^\star}\>{,}
\end{equation}
where $\Gamma_{\mathrm{v}}$ is the virtual photon flux
and ${\d}\sigma_{\mathrm{v}}/{\d}\Omega_\pi^\star$
is the virtual photon cross section in the CM frame
of the final $\pi\mathrm{N}$ system.  It can be further decomposed
into transverse, longitudinal and two interference parts,
\begin{equation}
{{\d}\sigma_{\mathrm{v}}\over{\d}\Omega_\pi^\star}=
{{\d}\sigma_{\mathrm{T}}\over{\d}\Omega_\pi^\star}+
\varepsilon_{\mathrm{L}}^\star\,
{{\d}\sigma_{\mathrm{L}}\over{\d}\Omega_\pi^\star}+
\sqrt{ 2\,\varepsilon_{\mathrm{L}}^\star(1+\varepsilon)}\,
{{\d}\sigma_{\mathrm{LT}}\over
{\d}\Omega_\pi^\star}\,\cos\phi_\pi +
\varepsilon {{\d}\sigma_{\mathrm{TT}}\over
{\d}\Omega_\pi^\star}\,\cos 2\phi_\pi
\end{equation}
with the transverse ($\varepsilon$) and longitudinal
($\varepsilon_{\mathrm{L}}^\star = Q^2\varepsilon/\omega^{\star 2}$) 
polarisations of the virtual photon fixed by the electron kinematics.  
The measured cross sections are listed in Table~\ref{tab:cross}.

\vspace{5mm}

\begin{table}[h]
\caption{Measured cross sections in the
$\mathrm{p}(\mathrm{e},\mathrm{e}'\pi^+)\mathrm{n}$ reaction.}
\label{tab:cross}
\begin{center}
\vspace*{6pt}
\begin{tabular}{crcc}
\hline

\hline

\hline
     Setting ($\varepsilon$)
   & ${\d}\sigma/{\d}\Omega_\pi^\star$
   & Stat. error
   & Syst. error \\
   & $[\,\mu\mathrm{b}/\mathrm{sr}\,]$
   & $[\,\mu\mathrm{b}/\mathrm{sr}\,]$
   & $[\,\mu\mathrm{b}/\mathrm{sr}\,]$ \\
\hline\\
$0.834$ & $11.14$  & $0.08$ ($0.7$\%) & $0.41$ ($3.7$\%) \\
$0.500$ & $8.40$   & $0.11$ ($1.3$\%) & $0.31$ ($3.7$\%) \\
$0.219$ & $5.96$   & $0.14$ ($2.3$\%) & $0.19$ ($3.2$\%) \\

$0.742$ & $7.48$   & $0.11$ ($1.5$\%) & $0.18$ ($2.4$\%) \\
$0.437$ & $5.36$   & $0.09$ ($1.7$\%) & $0.10$ ($1.8$\%) \\
$0.229$ & $4.55$   & $0.10$ ($2.1$\%) & $0.07$ ($1.6$\%) \\

$0.648$ & $5.03$   & $0.05$ ($1.1$\%) & $0.08$ ($1.6$\%) \\
$0.457$ & $4.19$   & $0.06$ ($1.3$\%) & $0.08$ ($1.8$\%) \\
$0.259$ & $3.46$   & $0.05$ ($1.4$\%) & $0.07$ ($1.9$\%) \\
[5pt]
\hline

\hline

\hline
\end{tabular}
\end{center}
\end{table}

\begin{figure}[ht]
\begin{center}
\includegraphics[height=8cm]{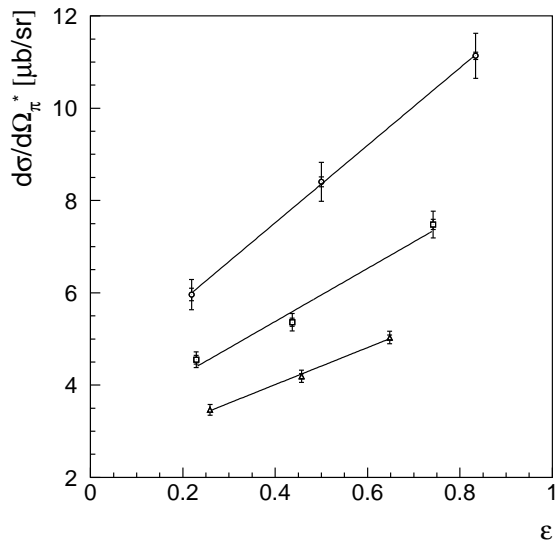}
\end{center}
\vspace*{-5mm}
\caption{Least-squares straight-line fits to the measured cross
sections for $\mathrm{p}(\mathrm{e},\mathrm{e}'\pi^+)\mathrm{n}$
at $W=1125\,\mathrm{MeV}$, for three values of $Q^2$.
The smaller error bars correspond to statistical, the larger ones
to the sum of statistical and systematical errors.}
\label{fig:3lines}
\end{figure}

\begin{figure}[ht]
\begin{center}
\includegraphics[height=7.5cm]{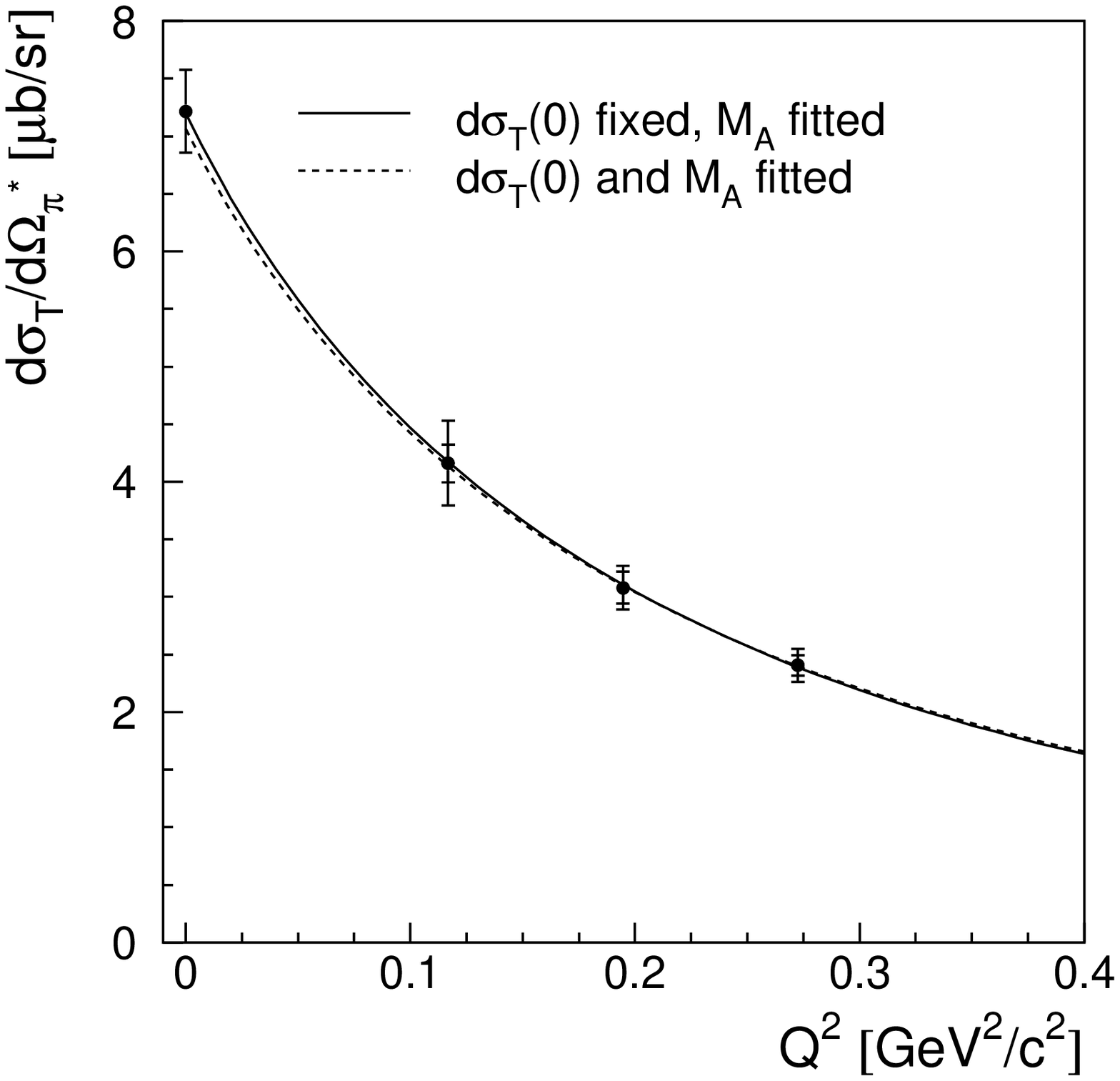}
\vspace*{-5mm}
\includegraphics[height=7.5cm]{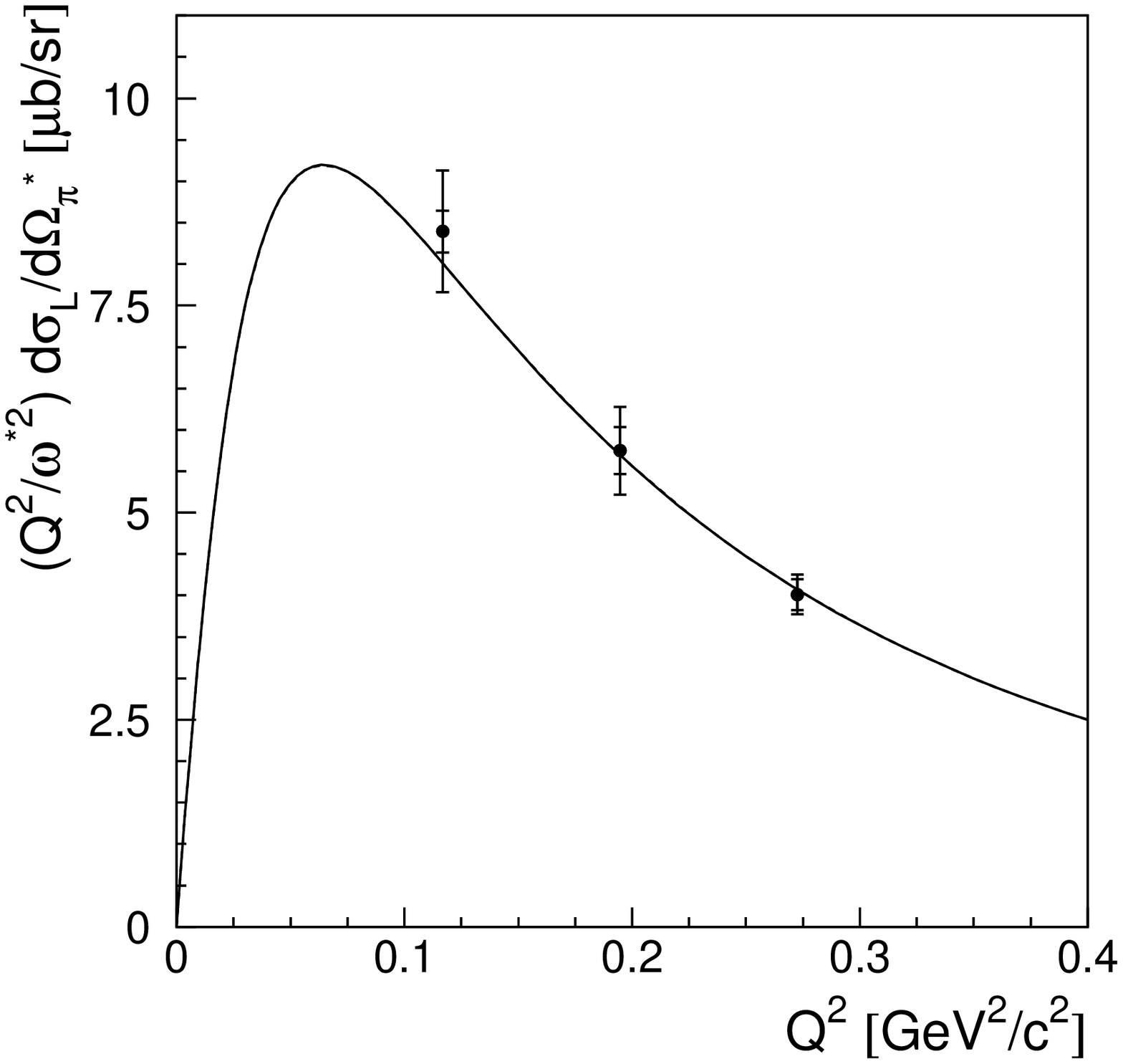}
\end{center}
\caption{Separated transverse and longitudinal cross sections.
The solid line shows our fit with
${\d}\sigma_{\mathrm{T}}/{\d}\Omega_\pi^\star(Q^2=0)$
fixed to $(7.22\pm 0.36)\,\mu\mathrm{b}/\mathrm{sr}$;
the dotted line is the unconstrained fit.  In the longitudinal
part the fits are almost indistinguishable.
The smaller error bars correspond to statistical, the larger ones
to the sum of statistical and systematical errors.}
\label{fig:wqs2}
\end{figure}

In parallel kinematics
($\theta_\pi^\star=\theta_\pi=0^\circ$) the interference parts
vanish due to their $\sin\theta_\pi^\star$ and
$\sin^2\theta_\pi^\star$ dependence.  At constant $Q^2$,
the transverse and the longitudinal cross sections
can therefore be separated using the Rosenbluth method
by varying $\varepsilon$,
\begin{equation}
{{\d}\sigma_{\mathrm{v}}\over{\d}\Omega_\pi^\star}=
{{\d}\sigma_{\mathrm{T}}\over
{\d}\Omega_\pi^\star}
+\varepsilon\,{Q^2\over\omega^{\star 2}}\,
{{\d}\sigma_{\mathrm{L}}\over
{\d}\Omega_\pi^\star}\>\mathrm{.}
\label{eq:rosen}
\end{equation}
Not only the statistical, but also the $\varepsilon$-correlated
systematical uncertainties of the data were considered
in our fit, whereas the $\varepsilon$-independent systematical
errors were included in the final uncertainty of
${\d}\sigma_{\mathrm{T}}$ and ${\d}\sigma_{\mathrm{L}}$.
The results are shown in Fig.~\ref{fig:3lines}
and Table~\ref{tab:LT}.  The $Q^2$-dependence of the separated
transverse and longitudinal cross sections can be seen
in Fig.~\ref{fig:wqs2}, together with the theoretical fits
used to extract the form factors.

\bigskip

\begin{table}[h]
\caption{The results of the L/T separation based on the
least-squares fit to the data.}
\label{tab:LT}
\begin{center}
\vspace*{6pt}
\begin{tabular}{ccc}
\hline

\hline

\hline
  Setting ($Q^2$)
& ${\d}\sigma_{\mathrm{T}}/{\d}\Omega_\pi^\star$
& $(Q^2/\omega^{\star 2})\,
   {\d}\sigma_{\mathrm{L}}/{\d}\Omega_\pi^\star$ \\
  $[\,\mathrm{GeV}^2/\mathrm{c}^2\,]$
& $[\,\mu\mathrm{b}/\mathrm{sr}\,]$
& $[\,\mu\mathrm{b}/\mathrm{sr}\,]$ \\
\hline\\
$0.117$ & $4.160\pm 0.165_\mathrm{stat}\pm 0.202_\mathrm{sys}$
        & $8.394\pm 0.254_\mathrm{stat}\pm 0.481_\mathrm{sys}$ \\
$0.195$ & $3.080\pm 0.139_\mathrm{stat}\pm 0.051_\mathrm{sys}$
        & $5.747\pm 0.284_\mathrm{stat}\pm 0.246_\mathrm{sys}$ \\
$0.273$ & $2.406\pm 0.088_\mathrm{stat}\pm 0.056_\mathrm{sys}$
        & $4.012\pm 0.187_\mathrm{stat}\pm 0.052_\mathrm{sys}$ \\
[5pt]
\hline

\hline

\hline
\end{tabular}
\end{center}
\end{table}

Since values of $\nu=-Q^2/M^2$ and $W$ in this experiment
were too high for a direct application of $\chi\mathrm{PT}$,
an effective Lagrangian model \cite{piplus1,dt} was used
to analyse the measured $Q^2$-dependence of the cross section,
and to extract the nucleon axial and pion
charge form factors.  In the energy region of our experiment,
the pseudovector $\pi\mathrm{NN}$ coupling evaluated at tree-level
provided an adequate description of the reaction cross section.
We included the $s$- and $u$-channel nucleon pole terms
containing electric and magnetic Sachs nucleon form factors
of the well-known dipole form with a `cut-off' 
$\Lambda_{\mathrm{d}}=0.843\,\mathrm{GeV}$,
the $t$-channel pion pole term with a monopole form factor
$F_\pi(Q^2)=1/(1+Q^2/\Lambda_\pi^2)$,
the contact (seagull) term with the axial dipole form factor
$G_{\mathrm{A}}(Q^2)=G_{\mathrm{A}}(0)/(1+Q^2/M_{\mathrm{A}}^2)^2$,
and the $s$-channel $\Delta$-resonance term.  Vector meson exchange
contributions in the $t$-channel were found to play
a negligible role in the charged pion channel.

Due to cancellations between higher partial waves and interference
terms with the $s$-wave, ${\d}\sigma_{\mathrm{T}}$ is
predominantly sensitive to the $E_{0+}(\mathrm{n}\pi^+)$ amplitude 
and therefore to $M_{\mathrm{A}}$.  On the other hand,
the pion charge form factor appears in the longitudinal
amplitude $L_{0+}(\mathrm{n}\pi^+)$ only at order 
$\mathcal{O}(\mu^2,\mu\nu)$, and the $s$-wave contribution to
${\d}\sigma_{\mathrm{L}}$ amounts to $10\,\%$ only.
Due to the contributions of the higher partial waves,
however, the longitudinal cross section ${\d}\sigma_{\mathrm{L}}$
is quite sensitive to $\Lambda_\pi$, which is a bonus `by-product'
of the analysis.

Since current conservation is violated if arbitrary form factors
are included, gauge invariance was imposed
by additional gauge terms in the hadronic current.
These terms modify the longitudinal part of the
cross section and therefore influence the pion
pole term and thus the extracted value of $\Lambda_\pi$. 
However, this procedure does not affect the transverse
part of the cross section.  The axial mass can then be
determined from the $Q^2$-dependence of ${\d}\sigma_{\mathrm{T}}$,
and the result can be compared to the prediction of $\chi\mathrm{PT}$.
In the theoretical fit, the transverse part is fitted first.

\newpage

We have used two different techniques to determine the axial mass:

(I) Only the axial mass was varied, whereas the value of
${\d}\sigma_{\mathrm{T}}$ at $Q^2=0$ was fixed by extrapolating
the transverse cross section (i.~e.~the $E_{0+}(\mathrm{n}\pi^+)$
amplitude) to the value of the photoproduction angular distribution
at $\theta_\pi=0^\circ$.  A precise cross section at the photon
point ($Q^2=0$) is very helpful for our analysis and can reduce
the uncertainty in determining the axial mass considerably.
Unfortunately, there are no experimental data available
around $W=1125\,\mathrm{MeV}$ and forward angles.
The only measurements in this energy region are performed
at pion angles larger than $60^\circ$ with large deviations
among the different data sets.
Therefore we have used a value at the photon point
obtained from different partial-wave analyses of the VPI group
\cite{said} and of the Mainz dispersion analysis \cite{hanstein}.
This yields to a weighted-average cross section at the photon point
of $(7.22\pm 0.36)\,\mu\mathrm{b}/\mathrm{sr}$
and this value was used as an additional data point.
The corresponding value of $E_{0+}(\mathrm{n}\pi^+)$ is also 
well supported by the studies of the GDH sum rule \cite{gdh}
and by the low energy theorem (Kroll-Ruderman limit).

(II) The three data points alone were fitted, while the value
of the transverse cross section at $Q^2=0$ was taken
as an additional parameter, with the result
${\d}\sigma_{\mathrm{T}}(0)=
(7.06\pm 1.12)\,\mu\mathrm{b}/\mathrm{sr}$.

The best-fit parameters for the transverse cross section
were then used to fit the longitudinal part.

Using the first and preferred procedure,
we find from the transverse cross section
$M_{\mathrm{A}}=(1.077\pm 0.039)\,\mathrm{GeV}$,
corresponding to
$\langle r^2_{\mathrm{A}}\rangle^{1/2}=
(0.635\pm 0.023)\,\mathrm{fm}$.
From the longitudinal part, we obtain
$\Lambda_\pi=(0.654\pm 0.027)\,\mathrm{GeV}$, corresponding to
$\langle r_{\pi}^2\rangle^{1/2}=(0.740\pm 0.031)\,\mathrm{fm}$.
The second procedure leads to the following results:
$M_{\mathrm{A}}=(1.089\pm 0.106)\,\mathrm{GeV}$ or
$\langle r^2_{\mathrm{A}}\rangle^{1/2}=
(0.628\pm 0.061)\,\mathrm{fm}$, and
$\Lambda_\pi=(0.658\pm 0.028)\,\mathrm{GeV}$ or
$\langle r_{\pi}^2\rangle^{1/2}=(0.734\pm 0.031)\,\mathrm{fm}$.

\section{Summary and conclusions}

We have measured the electroproduction of positive
pions on protons at the invariant mass of $W=1125\,\mathrm{MeV}$,
and at four-momentum transfers of $Q^2=0.195\,(\mathrm{GeV/c})^2$
and $0.273\,(\mathrm{GeV/c})^2$.
In conjunction with our previous measurement at
$Q^2=0.117\,(\mathrm{GeV/c})^2$ \cite{piplus1},
we were then able to study the $Q^2$-dependence
of the transverse and longitudinal cross sections,
separated by the Rosenbluth technique for each $Q^2$.
The statistical uncertainties were between $0.7\,\%$ and $2.3\,\%$,
an improvement of an order of magnitude over the result
of Ref.~\cite{bardin}.  The systematical uncertainties were
estimated to be between $1.6\,\%$ and $3.7\,\%$, and are expected
to decrease significantly in the future experiments.

We have extracted the axial mass parameter $M_{\mathrm{A}}$
of the nucleon axial form factor from our pion electroproduction
data using an effective Lagrangian model with pseudovector
$\pi\mathrm{NN}$ coupling.  Our extracted value of
$M_{\mathrm{A}}=(1.077\pm 0.039)\,\mathrm{GeV}$ is
$(0.051\pm 0.044)\,\mathrm{GeV}$ larger than the axial mass
$M_{\mathrm{A}}=(1.026\pm 0.021)\,\mathrm{GeV}$
known from neutrino scattering experiments. Our result
essentially confirms with the scaled-error weighted average
$M_{\mathrm{A}}=(1.068\pm 0.017)\,\mathrm{GeV}$
of older pion electroproduction experiments.
If we include our value into the database, the weighted average
increases to $M_{\mathrm{A}}=(1.069\pm 0.016)\,\mathrm{GeV}$,
and the `axial mass discrepancy' becomes
$\Delta M_{\mathrm{A}}=(0.043\pm 0.026)\,\mathrm{GeV}$.
This value of $\Delta M_{\mathrm{A}}$
is in agreement with the prediction derived from $\chi\mathrm{PT}$,
$\Delta M_{\mathrm{A}}=0.056\,\mathrm{GeV}$.
We conclude that the puzzle of seemingly different axial radii
as extracted from pion electroproduction and neutrino scattering
can be resolved by pion loop corrections to the former process,
and that the size of the predicted corrections is confirmed
by our experiment.

Theoretical input needed to extract the axial mass
and the pion radius has naturally led to some model dependence
of the results.  The dominant contribution to pion
electroproduction at $W=1125\,\mathrm{MeV}$ is due to the Born terms.
These are based on very fundamental grounds and the couplings
are very well known.  For our purpose the electric
and magnetic form factors of the nucleons are also accurately known
whereas the remaining two (the axial and the pion form factors)
are the subjects of our analysis.
In parallel kinematics we are in the ideal situation
where the pion form factor contributes only to the
longitudinal cross section, therefore reducing very strongly
the model dependence of the transverse cross section
and consequently of the determination of the axial mass.
Furthermore, the $\Delta$ resonance, which plays the
second important role in our theoretical description,
also contributes with a well-known M1 excitation to the transverse
cross section, while the longitudinal C2 excitation gives rise
to a larger uncertainty due to the less known C2 form factor,
currently under investigation at different laboratories.

Altogether, the model dependence is smaller for extracting
the axial mass than for the pion radius.
This could partly explain the discrepancy in the different values
of the pion radius between our analysis and the analysis of
pion scattering off atomic electrons \cite{amendo}.
However, as in the case of the axial mass,
an additional correction of the pion radius obtained from
electroproduction experiments is very likely.
This should be investigated in future studies
of chiral perturbation theory.


\begin{ack}
This work was supported by the Deutsche Forschungsgemeinschaft,
SFB 201.
\end{ack}



\newpage

\begin{thebibliography}{99}
\raggedright\frenchspacing
\bibitem{fanourakisX} G.~Fanourakis et al.,
    Phys. Rev. D {\bf 21} (1980) 562;\\
  L.~A.~Ahrens et al., Phys. Rev. D {\bf 35} (1987) 785;\\
  L.~A.~Ahrens et al., Phys. Lett. B {\bf 202} (1988) 284.
\bibitem{barishX} S.~J.~Barish et al.,
    Phys. Rev. D {\bf 16} (1977) 3103;\\
  K.~L.~Miller et al., Phys. Rev. D {\bf 26} (1982) 537;\\
  W.~A.~Mann et al., Phys. Rev. Lett. {\bf 31} (1973) 844;\\
  N.~J.~Baker et al., Phys. Rev. D {\bf 23} (1981) 2499;\\
  T.~Kitagaki et al., Phys. Rev. D {\bf 28} (1983) 436;\\
  T.~Kitagaki et al., Phys. Rev. D {\bf 42} (1990) 1331.
\bibitem{holder} M.~Holder et al., Nuovo Cim. A {\bf LVII} (1968) 338.
\bibitem{kustom} R.~L.~Kustom et al.,
  Phys. Rev. Lett. {\bf 22} (1969) 1014.
\bibitem{perkins} D.~Perkins, in: {\sl Proceedings of the\/}
  $16^{\mathrm{th}}$ {\sl International Conference
  on High Energy Physics\/}, J.~D.~Jackson, A.~Roberts (eds.),
  National Accelerator Laboratory, Batavia, Illinois, 1973,
  Vol.~IV, 189.
\bibitem{orkin} A.~Orkin-Lecourtois and C.~A.~Piketty,
  Nuovo Cim. A {\bf L} (1967) 927.
\bibitem{bonetti} S.~Bonetti et al.,
  Nuovo Cim. A {\bf 38} (1977) 260.
\bibitem{armenise} N.~Armenise et al.,
  Nucl. Phys. B {\bf 152} (1979) 365.
\bibitem{budagov} I.~Budagov et al.,
  Lett. Nuovo Cim. {\bf II} (1969) 689.
\bibitem{PDG} C.~Caso et al. (Particle Data Group),
  {\sl Review of Particle Properties\/},
  Eur. Phys. J. C {\bf 3} (1998) 9.
\bibitem{esau} A.~S.~Esaulov, A.~M.~Pilipenko, Yu.~I.~Titov,
  Nucl. Phys. B {\bf 136} (1978) 511.
\bibitem{olsson} M.~G.~Olsson, E.~T.~Osypowski and E.~H.~Monsay,
  Phys. Rev. D {\bf 17} (1978) 2938.
\bibitem{amaldiX} E.~Amaldi et al.,
    Nuovo Cim. A {\bf LXV} (1970) 377;\\
  E.~Amaldi et al., Phys. Lett. B {\bf 41} (1972) 216;\\
  P.~Brauel et al., Phys.~Lett.~B {\bf 45} (1973) 389;\\
  A.~del~Guerra et al., Nucl. Phys. B {\bf 99} (1975) 253;\\
  A.~del~Guerra et al., Nucl. Phys. B {\bf 107} (1976) 65.
\bibitem{joos} P.~Joos et al., Phys. Lett. B {\bf 62} (1976) 230.
\bibitem{choi} S.~Choi et al.,
  Phys. Rev. Lett. {\bf 71} (1993) 3927.
\bibitem{nambu} Y.~Nambu and M.~Yoshimura,
  Phys. Rev. Lett. {\bf 24} (1970) 25.
\bibitem{bloo} E.~D.~Bloom et al.,
  Phys. Rev. Lett. {\bf 30} (1973) 1186.
\bibitem{KR} N.~M.~Kroll and M.~A.~Ruderman,
  Phys. Rev. {\bf 93} (1954) 233.
\bibitem{NLS} Y.~Nambu and D.~Luri\'e,
    Phys. Rev. {\bf 125} (1962) 1429;\\
  Y.~Nambu and E.~Shrauner, Phys. Rev. {\bf 128} (1962) 862.
\bibitem{FPV} G.~Furlan, N.~Paver and C.~Verzegnassi,
  Nuovo Cim. A {\bf LXX} (1970) 247;\\
  C.~Verzegnassi,
    Springer Tracts in Modern Physics {\bf 59} (1971) 154;\\
  G.~Furlan, N.~Paver and C.~Verzegnassi,
  Springer Tracts in Modern Physics {\bf 62} (1972) 118.
\bibitem{DR} N.~Dombey and B.~J.~Read,
    Nucl. Phys. B {\bf 60} (1973) 65;\\
  B.~J.~Read, Nucl. Phys. B {\bf 74} (1974) 482.
\bibitem{BNR} G.~Benfatto, F.~Nicol\`o and G.~C.~Rossi,
  Nucl. Phys. B {\bf 50} (1972) 205;\\
  G.~Benfatto, F.~Nicol\`o and G.~C.~Rossi,
    Nuovo Cim. A {\bf 14} (1973) 425.
\bibitem{piplus1} K. I. Blomqvist et al. (A1 Collaboration),
  Z. Phys. A {\bf 353} (1996) 415.
\bibitem{dt}
  D.~Drechsel and L.~Tiator,
    J.~Phys.~G: Nucl. Part. Phys. {\bf 18} (1992) 449.
\bibitem{BKM1} V.~Bernard, N.~Kaiser and U.--G.~Mei\ss ner,
  Phys. Rev. Lett. {\bf 69} (1992) 1877;\\
  V.~Bernard, N.~Kaiser and U.--G.~Mei\ss ner,
  Phys. Rev. Lett. {\bf 72} (1994) 2810.
\bibitem{mami} H.~Herminghaus et al., Proc. LINAC Conf. 1990,
  Albuquerque, New Mexico;\\
  J.~Ahrens et al., Nucl. Phys. News {\bf 2} (1994) 5.
\bibitem{neuhaus} K.~I.~Blomqvist et al.,
  Nucl. Instr. Meth. A {\bf 403} (1998) 263.
\bibitem{xs} E.~Amaldi, S.~Fubini and G.~Furlan,
  Springer Tracts in Modern Physics {\bf 83} (1979) 6.
\bibitem{said} R.~A.~Arndt, I.~I.~Strakovsky and R.~L.~Workman,
  Phys. Rev. C {\bf 53} (1996) 430;\\
  see also the web-page {\tt http://said.phys.vt.edu/analysis/go3pr.html}.
\bibitem{hanstein} O.~Hanstein, D.~Drechsel, L.~Tiator,
  Nucl. Phys. A {\bf 632} (1998) 561.
\bibitem{gdh} D.~Drechsel and G.~Krein,
  Phys. Rev. D {\bf 58} (1998) 116009.
\bibitem{amendo} S.~R.~Amendolia et al.,
    Phys. Lett. B {\bf 146} (1984) 116;\\
  S.~R.~Amendolia et al.,
    Phys. Lett. B {\bf 178} (1986) 435.
\bibitem{bardin} G.~Bardin et al., Nucl. Phys. B {\bf 120} (1977) 45.
\end{thebibliography}
\end{document}